\documentclass[aps,pra,twocolumn,showpacs,floatfix]{revtex4-1}

	\usepackage{amsmath}    
	\usepackage{amsfonts}  
	\usepackage{amssymb}
\usepackage{epstopdf}
\usepackage{verbatim}
\usepackage{color}
\usepackage{appendix}
\usepackage[colorlinks, linkcolor=blue, citecolor=blue, urlcolor=blue,breaklinks=true]{hyperref}
\usepackage{soul}

	\usepackage{graphicx}   
	\usepackage{float}

\graphicspath{{figures/}}

\newcommand{\DeltaM}{\ensuremath{\Delta_{\rm M}}}
\newcommand{\DeltaA}{\ensuremath{\Delta_{\rm A}}}

\providecommand{\ket}[1]{\ensuremath{\left\vert #1\right\rangle}}

\providecommand{\comm}[2]{\left[#1,#2\right]}
\providecommand\avr[1]{\left\langle #1 \right\rangle}

\providecommand{\abs}[1]{\left\vert #1\right\vert}

\providecommand{\lp}{\left(}
\providecommand{\rp}{\right)}

\begin{document}
\title{Bistability effect in the extreme strong coupling regime of the Jaynes-Cummings model}
\author{A.~Dombi}
\affiliation{Institute for Solid State Physics and Optics, Wigner Research Centre, Hungarian Academy of Sciences, H-1525 Budapest P.O. Box 49, Hungary}
\author{A.~Vukics}
\affiliation{Institute for Solid State Physics and Optics, Wigner Research Centre, Hungarian Academy of Sciences, H-1525 Budapest P.O. Box 49, Hungary}
\author{P.~Domokos}
\affiliation{Institute for Solid State Physics and Optics, Wigner Research Centre, Hungarian Academy of Sciences, H-1525 Budapest P.O. Box 49, Hungary}

\begin{abstract}We study the nonlinear response of a driven cavity QED system in the extreme strong coupling regime where the saturation photon number is below one by many orders of magnitude. In this regime, multi-photon resonances within the Jaynes--Cummings spectrum up to high order can  be resolved.  We identify an intensity and frequency range of the external coherent drive for which the system exhibits bistability instead of resonant multi-photon transitions.  The cavity field evolves into a  mixture of the vacuum and another quasi-classical state well separated in phase space. The corresponding time evolution of the outgoing intensity is a telegraph signal alternating between two attractors.  
\end{abstract}

\maketitle

\section{Introduction}
\label{intro}

Optical bistability is a benchmark of nonlinear light-matter interaction. Initially, many decades ago, it has been demonstrated for a macroscopic nonlinear medium, typically a saturable absorber or a Kerr-type dispersive medium, in a Fabry--P\'erot-type optical resonator \cite{Lugiato1984II}.  The effect consists in the multi-valued solution and hysteresis in the transmitted output mean-field intensity for a certain range of the input power and frequency. Subsequently, owing to the development of high-finesse optical microresonators, the bistability effect could be observed at much lower light excitation level, with the medium size also reduced to hundreds of atoms \cite{Rempe1991Optical,Sauer2004Cavity,Brecha1999N}.  Optical bistability can be theoretically described in the frame of a semiclassical mean-field approach based on the lossy and driven Jaynes--Cummings model.

With the advent of the strong coupling regime of cavity quantum electrodynamics (QED), however, a much more refined picture of the non-linearity in the matter-light coupling must be conceived. The figure of merit is the saturation photon number $n_{\rm sat}$ which defines that intracavity intensity where the atomic response to excitation becomes nonlinear according to the classical theory. Today, cavity QED extends to the range of  $n_{\rm sat} < 1$ which indicates an obviously quantum regime in the light-matter interaction.  Interestingly, some remnants of semiclassical bistability could be observed in a strong-coupling cavity QED system with only a few degrees of freedom and operated in the regime of $n_{\rm sat} \lesssim 1$ \cite{Armen2006Lowlying,Kerckhoff2011Remnants}.  However, more generally,  the nonlinear input-output relation is manifested beyond the mean-field level. Dramatic effects can occur in the quantum statistics of the output field intensity and it can give insight directly to the anharmonic energy spectrum of the coupled atom-field system  \cite{Schuster2008Nonlinear}.  Provided the low-lying excitation levels are spectrally resolved, various quantum applications can be envisaged. A prominent example is the photon blockade \cite{imamoglu1997strongly}, i.e., when a single-photon pulse can be transmitted through the cavity but not a two-photon or higher one \cite{birnbaum2005photon,Kubanek2008Twophoton,Lang2011Observation}. Another useful application is squeezed-light generation by a single atom \cite{ourjoumtsev2011observation}. Ultimately, stationary photon number states can also be prepared in the cavity \cite{Chough2000Single,Pellizzari1994Preparation}. 

Microwave circuit QED systems reached an unprecedented strong coupling regime of cavity QED, with saturation photon number  $n_{\rm sat} \ll 1$. The ratio of $g$ (coupling parameter between a single mode of the stripline resonator and the artificial atom) to the loss rates is far larger than in atomic cavity QED realizations: typically $\gamma, \kappa \lesssim g/100$,  where $\kappa$ is the cavity mode linewidth, and $\gamma$ is the characteristic decay rate of the electronic dipole system \cite{blais2004cavity}.  This is the regime which we refer to as `extreme' strong in the title. Here, the formal semiclassical mean-field solution is expected to be invalid in general. An exception is  the special case  of very large detuning  between the mode and the resonance of the artificial atom \cite{Boissonneault2008Nonlinear} since  the large detuning reduces the effective coupling between the two quantum systems. In this case, dispersive bistability can be expected and interpreted semiclassically \cite{bishop2010response,Boissonneault2010Improved,Peano2010Dynamical}, which is used, for example, for high-fidelity qubit readout \cite{Reed2010High,Englert2010Mesoscopic}.

In this paper,  we explore the nonlinear input-output relation of a resonator-driven circuit or cavity QED system in a broad frequency and intensity range of the driving. The resonator mode and the artificial atomic systems are assumed to be resonant. Two distinct frequency domains of the driving field can be identified: (1) For large detuning from the resonator mode, resolved multi-photon transitions in the low-excitation part of the Jaynes--Cummings ladder appear; (2) By tuning the external drive closer to resonance with the mode, the system evolves into a bistability-like steady-state. It is a mixture of two `semiclassical' states, represented by a two-peaked  Wigner quasi-distribution function in phase space. This is an unexpected result since the robust semiclassical bistable state is generated by a single atom. This solution is not connected by any limiting procedure to the result that could be obtained from an \emph{ab initio} semiclassical description. Moreover, the bimodal Wigner function is not present in the case of exact resonance, as opposed to the familiar case of semiclassical absorptive bistability.

The paper is structured as follows. In Section \ref{sec:Model}, we introduce the model and present the semiclassical mean-field solution for later reference. Section \ref{sec:NonlinearSpectrum} is devoted to the nonlinear features appearing in the mean-field cavity transmission as the pump intensity and frequency are tuned.  We show that for pump intensities beyond the linear response regime, multi-photon resonances appear in contrast to the semiclassically expected dispersive bistability.  As the main result of this paper, in Section \ref{sec:QuantumBistability}, we present a bistability solution emerging in a frequency range where the multi-photon transitions are of very high order and cannot be excited at the given pump intensity. We present the phase-space distribution of the steady state, which clearly manifests the mixture of states with confined fluctuations both in intensity and phase.  We also give a time-resolved picture which exhibits a telegraph signal of the average intensity of the field. Section \ref{sec:Conclusions} consists of our conclusions.

\section{Model}
\label{sec:Model}

\begin{figure}
	\begin{center}
		\includegraphics[width=0.98\linewidth]{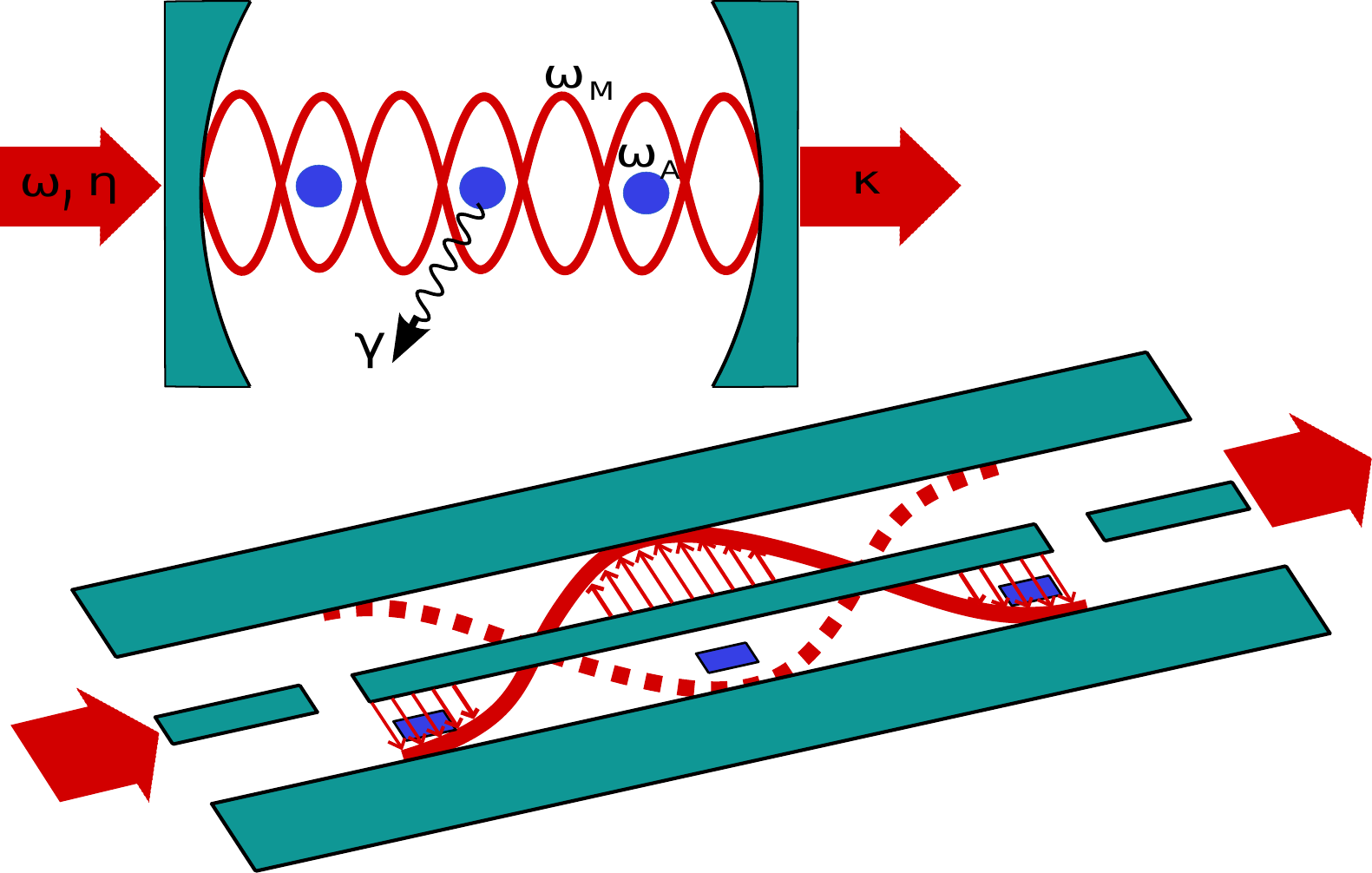}
	\end{center}
	\caption{\label{fig:Scheme} Schematic representation of the driven cavity QED system. The upper panel shows an atomic cavity QED system, i.e., atoms in a driven Fabry--P\'erot-type cavity. The input field is the pump with effective amplitude $\eta$ and frequency $\omega$, the output field is generated by the field leaking out from the cavity at a rate 2$\kappa$. The bottom panel represents the microwave circuit QED realization of the same Jaynes--Cummings model system. The single-mode field of the stripline resonator is coupled in and out by capacitive coupling to the transmission lines separated by the gaps in the middle line. The mode is resonant with artificial atoms that can be considered two-level electric dipole systems. }
\end{figure}

We consider a fixed number $N$ of identical two-level systems (atoms or artificial atoms) with resonance frequency $\omega_A$ coupled to a single mode of a high-finesse resonator with frequency $\omega_M$, as schematically shown in Fig.~\ref{fig:Scheme}.
The coupling to the mode is assumed to be uniform with strength $g$. The cavity is coherently driven with amplitude $\eta$ at a pump frequency $\omega$, the detunings from the cavity mode and from the atoms are denoted by $\DeltaM=\omega-\omega_{\rm M}$ and $\DeltaA=\omega-\omega_{\rm A}$, respectively. The system can be described by the Jaynes--Cummings Hamiltonian, which
reads ($\hbar=1$)
\begin{multline}
\label{eq:H}
H=-\DeltaM\,a^\dagger a - \DeltaA \sum_{i=1}^{N}\sigma^\dagger_i\,\sigma_i + i g \sum_{i=1}^{N} \lp a^\dagger\,\sigma_i-\sigma^\dagger_i\,a\rp \\ + i \eta\lp a^\dagger-a\rp.
\end{multline}
The bosonic annihilation and creation operators $a$ and $a^\dagger$ describe the radiation field mode, $\sigma_i$ and $\sigma^\dagger_i$ denote the lowering and raising operators for the two-level systems (indexed by $i=1\ldots N$). These latter complemented by the population inversion $\sigma_{z,i} = \sigma_i^\dagger \sigma_i - \frac{1}{2}$ and the unit operator form a complete set.  The algebra is equivalent to that of Pauli operators of a spin-$\frac{1}{2}$ particle, and we will refer to the atoms or artificial atoms as qubits. 

The system is dissipative and couples to the environment through several channels. We have cavity-photon loss due mainly to the photon outcoupling to propagating modes, and characterized by a rate $2 \kappa$. The qubit, in general, can have population and polarization damping with rates $\gamma_{\parallel}$ and $\gamma_{\perp}$, respectively. We assume zero-temperature reservoirs.  The corresponding Master equation in Lindblad form reads
\begin{multline}
\label{eq:Master}
\dot{\rho} = -i\comm{H}{\rho} + \kappa\lp 2 a \rho a^\dagger - a^\dagger a \rho - \rho a^\dagger a \rp \\ + \gamma_{\parallel}\sum_{i=1}^{N}\lp 2 \sigma_i \rho \sigma^\dagger_i - \sigma^\dagger_i\sigma_i \rho - \rho \sigma^\dagger_i\sigma_i  \rp \\ + 2 \gamma_{\perp}\sum_{i=1}^{N}\lp \sigma_{zi} \rho \sigma_{zi} - \frac{\rho}{4} \rp.
\end{multline}

Solid-state realizations of coupled matter and radiation modes give an experimental context for this model. One such system is in circuit-QED \cite{blais2004cavity}, where it is possible to couple several qubits to a single microwave cavity mode. The number of qubits can be varied by selectively detuning the resonance frequencies of certain qubits from the mode frequency \cite{fink2009dressed}. The parameters of our quantum simulations were modeled after one circuit-QED realization \cite{nissen2013collective}, where the single qubit-photon couplings of three superconducting transmon qubits were ${g} \simeq  2 \pi  \times 55$ MHz, the cavity decay rate is  ${\kappa} =  2 \pi  \times 0.47$ MHz, and for the qubit damping rates $\gamma_\parallel , {\gamma_\perp} \lesssim \kappa$.

It is important to note that the decay of the two-level systems is considered individually for each qubit, eluding the collective decay effect observed in Ref.~\cite{nissen2013collective}. Although the coupling of the qubits to the cavity mode is symmetric, the qubit ensemble cannot be replaced by a large collective spin because the spin decay will drive the system out of the subspace of states symmetric under permutation of the atoms. 

Concerning the calculation method,  we will unravel the full quantum Master equation defined by Eqs.~(\ref{eq:H}) and (\ref{eq:Master}) into quantum trajectories by means of the Monte Carlo wave-function method \cite{Dalibard1992Wave}.  The trajectories can be perceived as single experimental realizations of the dynamics and steady-state averages can be calculated by time averaging over a long trajectory. The actual simulations were performed using the C++QED framework \cite{VukicsCppQEDa,VukicsCppQEDb,Sandner2014CppQED}, which is a generic open-source C++/Python application-programming framework for efficient simulations of open quantum dynamics of interacting quantum system. 
 
The results of the full quantum calculation can be confronted with those originating from an \emph{ab initio} semiclassical mean-field model which we briefly recall here. Upon introducing the total spin operators $\Sigma=\sum_{i=1}^N\sigma_i$, $\Sigma_z=\sum_{i=1}^N\sigma_{z,i}$, a closed system of equations is obtained for the scaled mean field variables $\alpha=\avr a /\sqrt{N}$, $S=\avr\Sigma /N$, and $S_z=\avr{\Sigma_z}/N$, which reads
\begin{eqnarray}
\dot\alpha &=&\lp i\DeltaM-\kappa\rp\alpha + \sqrt{N} g\, S + \frac{\eta}{\sqrt{N}}, \nonumber\\
\dot S &=&\lp i\DeltaA-\gamma_{\perp}-\gamma_{\parallel}\rp S + 2 \sqrt{N} g\,S_z\,\alpha, \nonumber\\
\dot S_z &=&-2\gamma_{\parallel}\lp S_{z}+\frac{1}{2}\rp - \sqrt{N}g\lp S^*\,\alpha + \alpha^*\,S\rp. \label{eq:MeanField}
\end{eqnarray}
The steady-state solution of these equations for the intracavity intensity can be written as
\begin{multline}
\label{eq:Semiclassical}
\frac{|\eta|^2}{N} = | \alpha|^2 \, \Biggl( \left[\Delta_M-\Delta_A \, \frac{Ng^2}{\Delta_A^2 + \gamma_\perp^2\left(1 + |\alpha|^2/n_{\rm sat}\right)} \right]^2  \\ + 
\left[\kappa + \gamma_\perp \, \frac{Ng^2}{\Delta_A^2 + \gamma_\perp^2\left(1 + |\alpha|^2/n_{\rm sat}\right)} \right]^2  \Biggr).
\end{multline} 
This is an implicit equation for the mean intracavity intensity $|\alpha|^2$ as a function of the pump intensity, proportional to $|\eta|^2$, and frequency, included in $\Delta_M$ and $\Delta_A$.  
We will refer to this solution as the semiclassical one. It clearly manifests that the nonlinearity scales with the saturation photon number $n_{\rm sat} = \frac{\gamma_\parallel \gamma_\perp}{4 Ng^2}$. On the one hand, for intensities well below the saturation photon number, $|\alpha|^2 \ll n_{\rm sat}$, the spectrum of two coupled oscillators is recovered with a linear input-output relation ($|\alpha|^2 \propto |\eta|^2$). On the other hand, if $|\alpha|^2\gg n_{\rm sat}$, then the transmitted power tends to that of an empty resonator with the same Lorentzian spectrum. 

The semiclassical solution (\ref{eq:Semiclassical}) has a simple scaling for the number of atoms $N$. The solution for $ | \alpha|^2$ remains invariant if $Ng^2$ and $|\eta|^2/N$ is kept constant. That is, the same steady-state solution can be obtained with a single atom and large coupling strength as that for a large atomic ensemble with small coupling $g$.  It has been shown that a low number of atoms in the range of 5-8 suffices to well reproduce the semiclassical phase transition of absorptive bistability \cite{dombi2013optical}, which is a phase transition in the corresponding macroscopic system. In this paper, we will restrict the quantum calculations to the case of $N=1$.

An interesting regime occurs in the very strong coupling regime where $n_{\rm sat} \ll 1$, which allows for preserving the nonlinearity of the above equation with a mean intensity below or around $|\alpha|^2 \lesssim 1$. Such a tiny value is expected to invalidate the semiclassical description and the solution in Eq.~(\ref{eq:Semiclassical}). It is exactly this regime we will study in the following.

\section{From Rabi splitting to multi-photon resonance peaks}
\label{sec:NonlinearSpectrum}

Atomic cavity QED systems at present \cite{Sames2014Antiresonance} can realize couplings $\kappa \sim g/10$. In circuit QED, the regime $\kappa, \gamma_\perp, \gamma_\parallel \lesssim g/100$ is reached. For both cases, the saturation photon number is many orders of magnitude below 1. Specifically, for the parameter set characteristic of the experiment \cite{nissen2013collective},  the saturation photon number amounts to $n_{\rm sat} =6\times 10^{-7}$ for a single atom.  We consider a single atom contained in the cavity mode. 

\begin{figure}
			\raggedright \scriptsize{(a)} \\ 
			\includegraphics[width=0.90\linewidth]{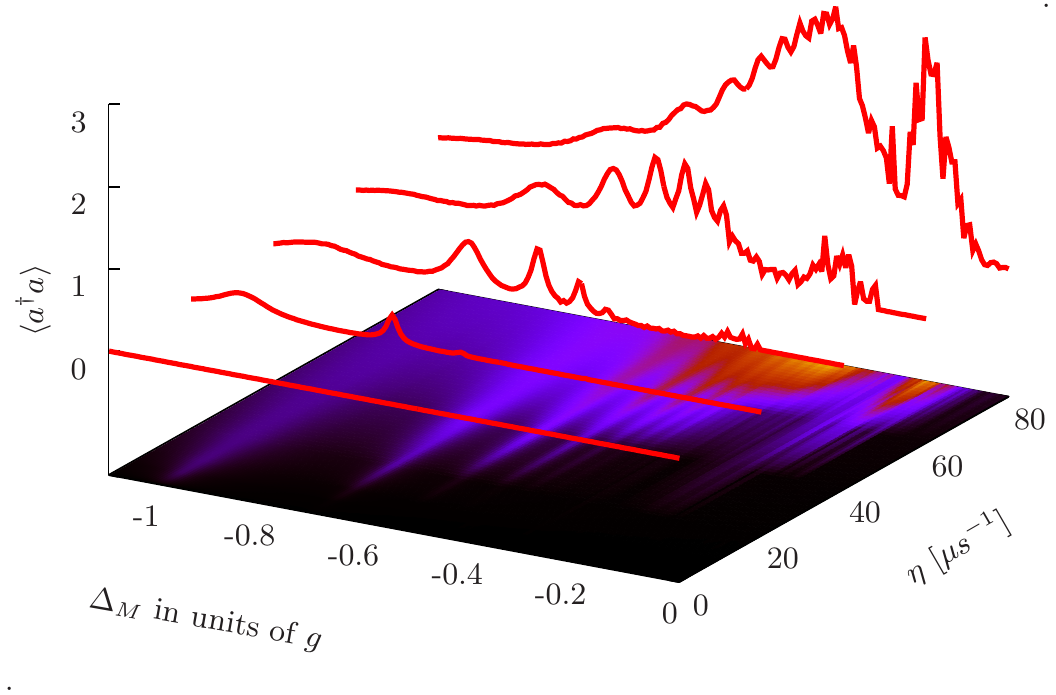} \\
			 \includegraphics[width=0.98\linewidth]{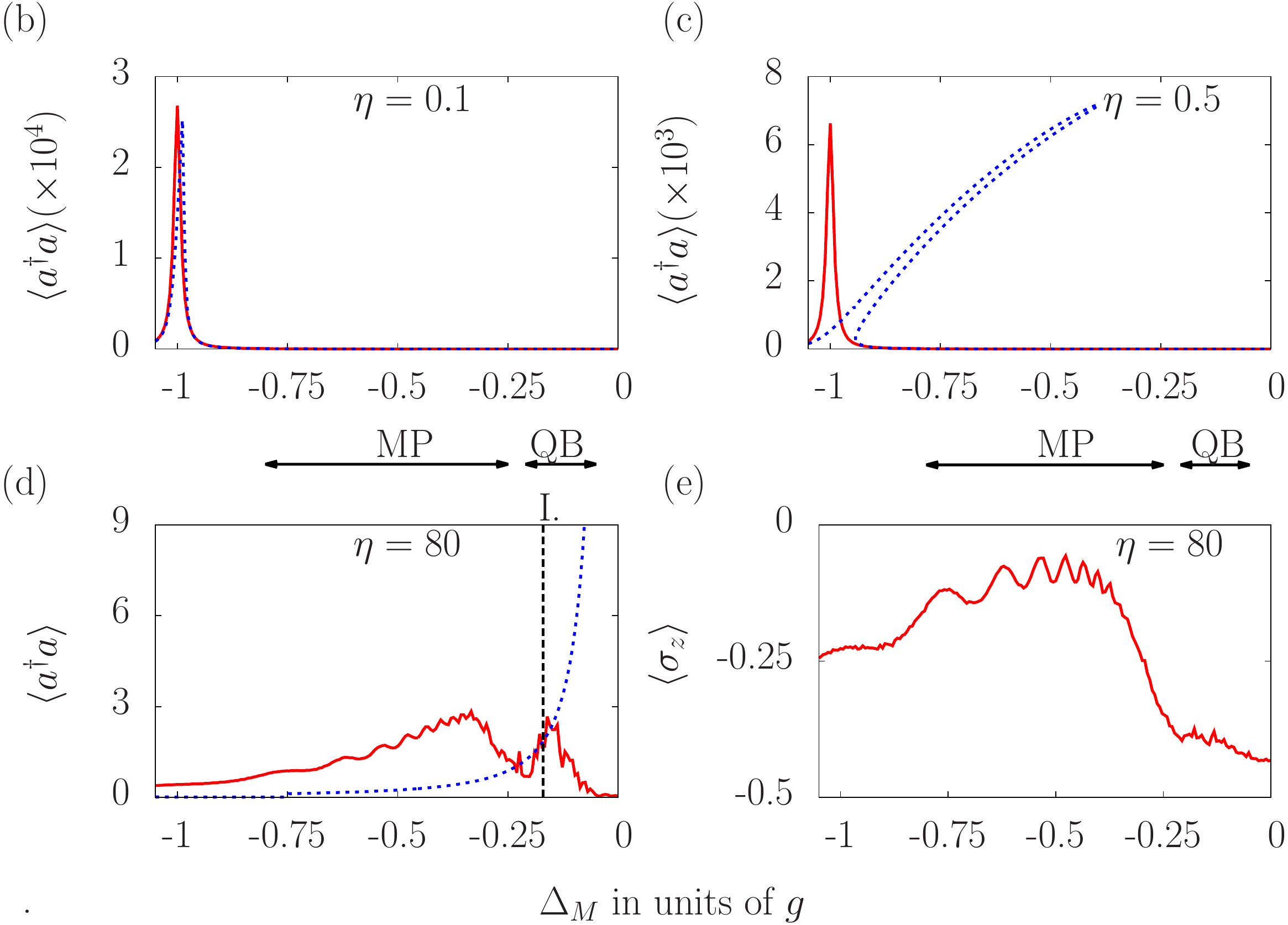} \\
	\caption{(a) Steady-state mean photon number as a function of the pump laser frequency for various driving strengths in the driven Jaynes-Cummings model (single atom $N=1$). We see a dramatic change of this spectrum upon increasing the pump power from the linear response (for $\eta=0.1 \mu$s${}^{-1}$, magnified in panel (b), and for $\eta=0.5 \mu$s${}^{-1}$, panel (c)), to multiple resolved resonance peaks corresponding to multi-photon transitions (for $\eta=80 \mu$s${}^{-1}$, panel (d), MP indicates schematically the domain of multi-photon resonances). The additional peak outside the multi-photon resonance structure close to $\Delta_M\approx -0.15 g$ is due to the quantum bistability effect (this domain is indicated by QB in the panels (d) and (e)). Dashed blue line represents the semiclassical solution. Panel (e) shows the atomic saturation exhibiting a very low value in the significant regime $-0.2\,g<\Delta_M$, at variance with the semiclassical expectation. Loss rates are  $\gamma_{\parallel}=2.95 \mu$s${}^{-1}$, $\gamma_{\perp}=0.1 \mu$s${}^{-1}$, $\kappa=2.95 \mu$s${}^{-1}$, the coupling strength is $g= 348 \mu$s${}^{-1}$. Atoms and the mode are resonant: $\omega_A=\omega_M$, which implies $\Delta_A=\Delta_M$.}
	\label{fig:spectrum_N1}
\end{figure}

We begin with surveying the mean-field manifestations of the nonlinear response of a driven cavity QED system in this extreme strong coupling regime. To this end, we present the mean intracavity intensity $\langle a^\dagger a\rangle$ as a function of the frequency of the monochromatic driving for a large range of driving intensity (or amplitude).  We will refer to this as \emph{transmission spectrum}, although in fact, the field intensity coupled out from the cavity and directed onto the photodetector is proportional to the intracavity photon number with a transmission coefficient which, in order to simplify the notation, we omit here\footnote{Note that the cavity loss rate $2\kappa$, in general, may involve other loss mechanisms than outcoupling, such as diffraction, absorption in the mirror, etc.}.   We assume resonance between the atoms and the mode  ($\omega_M=\omega_A$), hence the pump frequency can be expressed simply by the detuning $\Delta_M = \omega -\omega_M$.  Transmission spectra as a function of $\Delta_M$ are presented in Figure~\ref{fig:spectrum_N1}(a) for various selected values of the driving amplitude $\eta$. The dramatic variation of the spectrum lines when scanning  the pump amplitude over three orders of magnitude ($\eta =0.1 \ldots 80$) clearly evidences a strong non-linearity on such a large pump intensity scale (the intensity varies then six orders of magnitude).

For sufficiently low excitation level, the cavity QED system can be seen as a system of coupled linear oscillators. This is represented by the first spectrum line in the figure corresponding to $\eta=0.1 \mu$s${}^{-1}$. For a linear system, this spectrum is just the resolved vacuum Rabi splitting (only the negative detuning part is shown) which is the defining trait of strong coupling in the single atom\,--\,single cavity mode system. The expected spectrum is two Lorentzian peaks centered at $\Delta_M=\pm  g$. In this case, there must be a good agreement between the  semiclassical  and quantum descriptions even if the system is in the strong coupling regime of cavity QED.  This agreement is verified in Fig.~\ref{fig:spectrum_N1}(b) by the closely overlapping  solid red (quantum) and  dashed blue (semiclassical) curves, both being closely Lorentzian. 

When the drive amplitude is increased by a factor of 5, cf. Fig.~\ref{fig:spectrum_N1}(c), the quantum Monte Carlo wave function calculation leads to a transmission spectrum which retains the Lorentzian shape with 25-fold higher peak intensity. This result reflects a system still in the linear operation regime. At variance, the \emph{ab initio} semiclassical solution is a strongly distorted peak with multiple solutions in a given frequency range. In other words, the semiclassical description predicts a dispersive bistability at this driving strength for the given set of system parameters. For weak coupling cavity QED systems with many randomly positioned atoms, such a dispersive bistability has been verified experimentally \cite{Gripp1996Anharmonicity}. It is not surprising to find a significantly different behaviour in this extremely strong coupling regime. The photon number is in the range of $10^{-3}$ (cf. Figure), where the validity of the semiclassical description is not granted. 

\begin{figure}
	\begin{center}
		\includegraphics[width=0.5\linewidth]{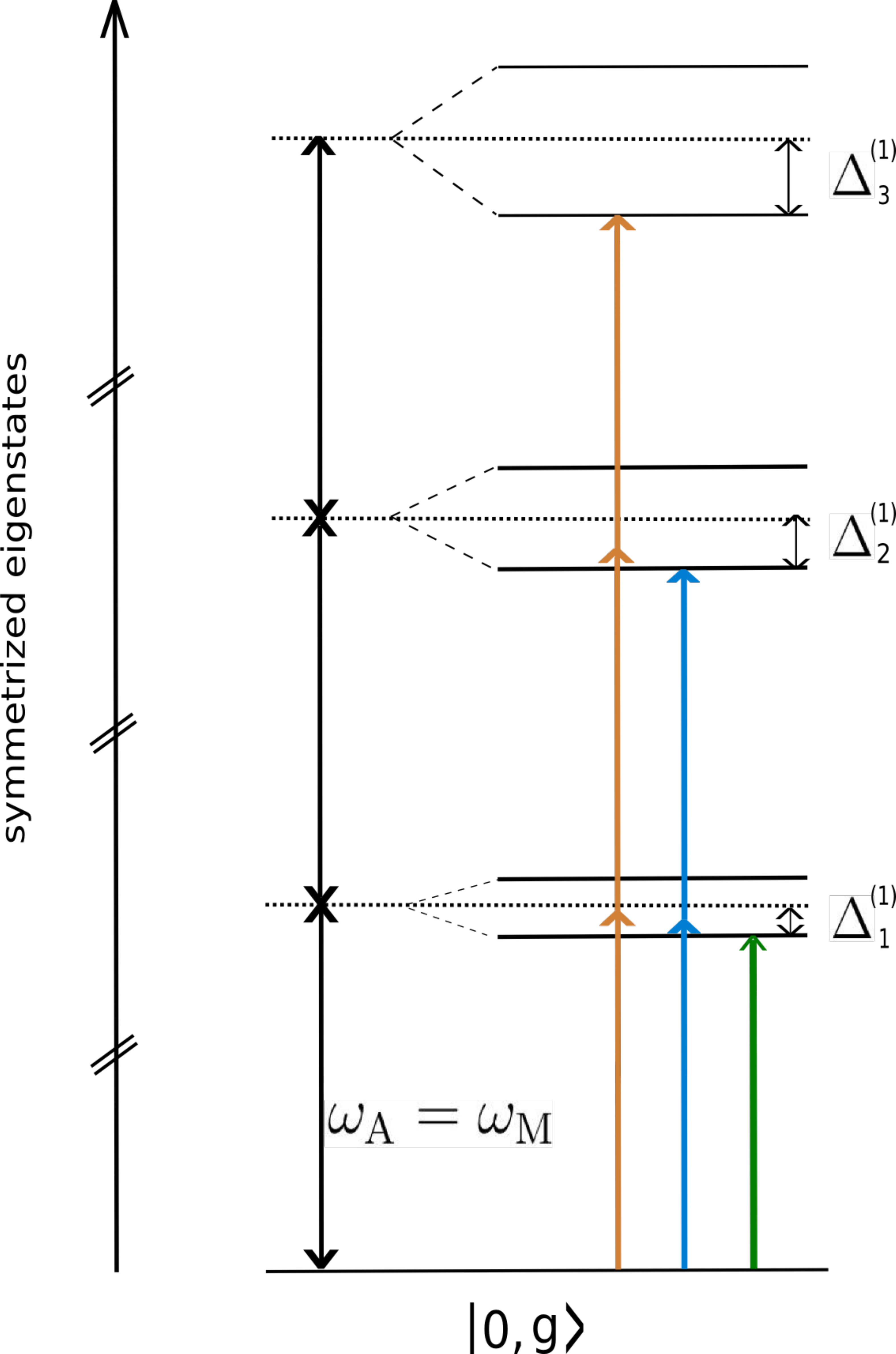}
	\end{center}
	\caption{\label{fig:JaynesCummings} Schematic representation of the multi-photon transitions in the single-atom Jaynes--Cummings model. The two-level systems and the boson mode are in resonance ($\omega_A=\omega_M$), which is the case throughout this paper.  Green, blue and orange arrows mark the expected one-, two-, and three-photon transitions, respectively.}
\end{figure}

The transmission spectrum for the next higher driving strength $\eta= 20 {\mu}s^{-1}$ exhibits two peaks, cf. Fig.~\ref{fig:spectrum_N1}(a). The one corresponding to the Rabi 
split resonance at $\Delta_M=- g$ is significantly broadened. In fact, this is the photon blockade regime, where the stronger pumping does not amount to more transmission. The eigenstates involving two or more photon numbers are far detuned due to the nonlinearity of the Jaynes--Cummings ladder. The system is effectively confined into the Hilbert space spanned by the ground state $|g,0\rangle$ and the lowest dressed state $|-,0\rangle$. The driven two-level system is saturated, hence the broadening of this peak. At the same time, a new peak appears at $\Delta_M=- g/\sqrt{2}$ corresponding to the two-photon transition to the higher dressed state $|-,1\rangle$. The low-lying multi-photon transitions  in the Jaynes--Cummings ladder are schematically shown in Fig.~\ref{fig:JaynesCummings}.

The next spectrum lines corresponding to $\eta = 40$, $60$, $80$  ${\mu}s^{-1}$, respectively, show the appearance of resolved resonances corresponding to multi-photon transitions to  higher and higher lying dressed states, see Fig.~\ref{fig:JaynesCummings}.  Such frequency dependence of the mean intensity was presented for the strong coupling regime in Ref.~\cite{shamailov2010multi} and have been observed experimentally  in circuit QED systems \cite{deppe2008two,bishop2009nonlinear,Kockum2013Deatiled}.  The resonance frequencies reflect the anharmonic Jaynes--Cummings spectrum \cite{fink2008climbing}-\footnote{For the sake of accuracy, we note that the resonance frequencies are characteristic of the driven rather than the bare Jaynes--Cummings  eigenfrequencies. The difference is the driving term proportional to $\eta$ in the Hamiltonian in Eq.~(\ref{eq:H}). The driving leads to a shift of the eigenfrequencies, which was calculated analytically in the special case of $\Delta_M=0$ and single atom \cite{alsing1992dynamic,Peano2010Quasienergy}.} 
 Similarly to the case of the one-photon transition, all the higher-order resonance lines undergo power broadening. Therefore, the resonances tend to merge into a broad structure, as can be clearly seen from the surface plots in Fig.~\ref{fig:spectrum_N1}(a). The multi-photon transitions arise from the strong driving of an effectively two-level system, only the decay processes, of course,  lead out from the two-state subspace.  

The multi-photon resonances are visible up to 10-photon transitions in this structure for the largest driving strength $\eta=80 \mu$s${}^{-1}$ considered, shown also separately in Fig.~\ref{fig:spectrum_N1}(d). When the detuning is decreased below $|\Delta_M|\lesssim0.35 g$ the mean transmitted intensity is reduced. In this small detuning range, resonance could occur only with very high-order transitions, for which the driving intensity $\eta=80 \mu$s${}^{-1}$ is not strong enough. This cutoff limits the multi-photon resonance regime which is indicated schematically by MP in Fig.~\ref{fig:spectrum_N1}(d).

It can be thus well understood that the multi-photon resonance peaks are suppressed as the detuning $\Delta_M$ tends to 0. Therefore it comes as a surprise that another significant peak emerges in the detuning range $-0.2 g < \Delta_M < -0.1 g$, indicated by QB in Fig.~\ref{fig:spectrum_N1}(d). The main result of this paper is the observation of this peak and its interpretation as a quantum bistability (QB) effect.

Let us first discuss if the presence of this peak can be explained by the semiclassical approach. On the rising slope of the peak ($-0.2 g \lesssim \Delta_M$), cf. Fig.~\ref{fig:spectrum_N1}(d), the numerical points fit nicely to the dashed blue curve representing the semiclassical solution. It can be immediately clarified that this fit does not signify any justification of the semiclassical picture.  The bistability regime according to the solution Eq.~(\ref{eq:Semiclassical}) is limited to the range $\eta \leq 20 \mu$s${}^{-1}$. For such a strong driving as we have here ($\eta=80 \mu$s${}^{-1}$), the atomic system inside the cavity  would be completely saturated within the semiclassical picture and the transmission would be that of an empty cavity (no atoms inside).  As a sharp distinction, the saturation in the quantum model is found to be very low in the QB domain, as shown in panel (e) of Fig.~\ref{fig:spectrum_N1}. In the next section, we will see that the photon number distribution is also significantly different from that of a driven empty cavity. To this end, we need to go beyond studying only the mean intensity of the mode. 

\section{Quantum bistability}
\label{sec:QuantumBistability}

\begin{figure}
	\begin{center}
		 (a)\\ \includegraphics[width=0.45\linewidth]{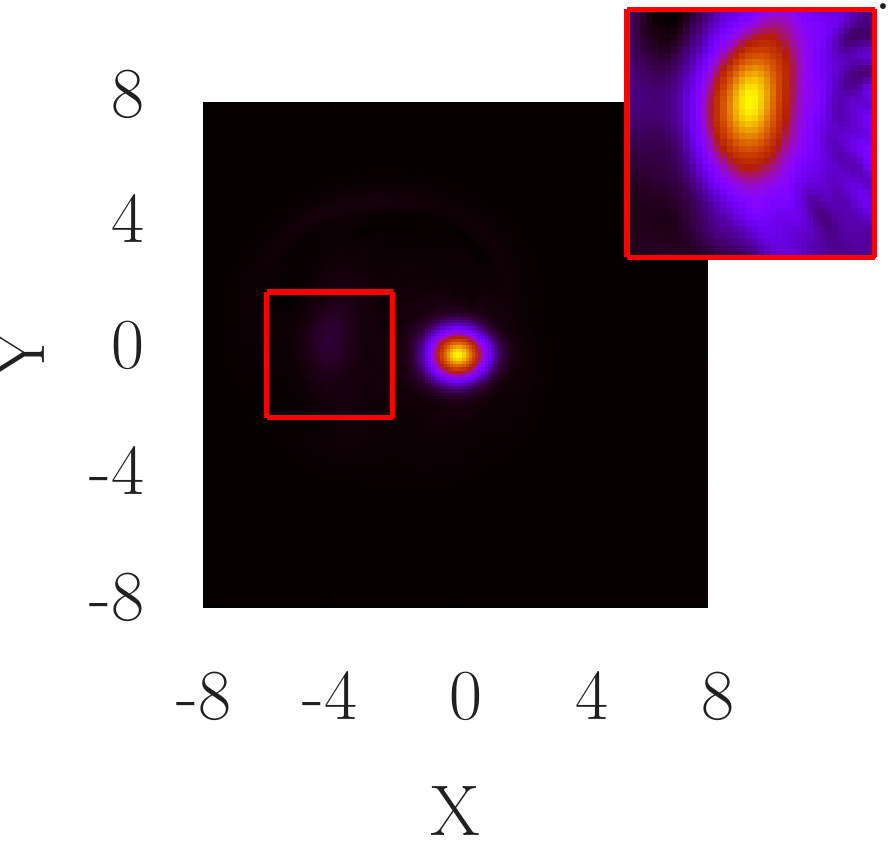}\\
		 \begin{tabular}{c c}
		 (b) & (c) \\
		 \includegraphics[width=0.45\linewidth]{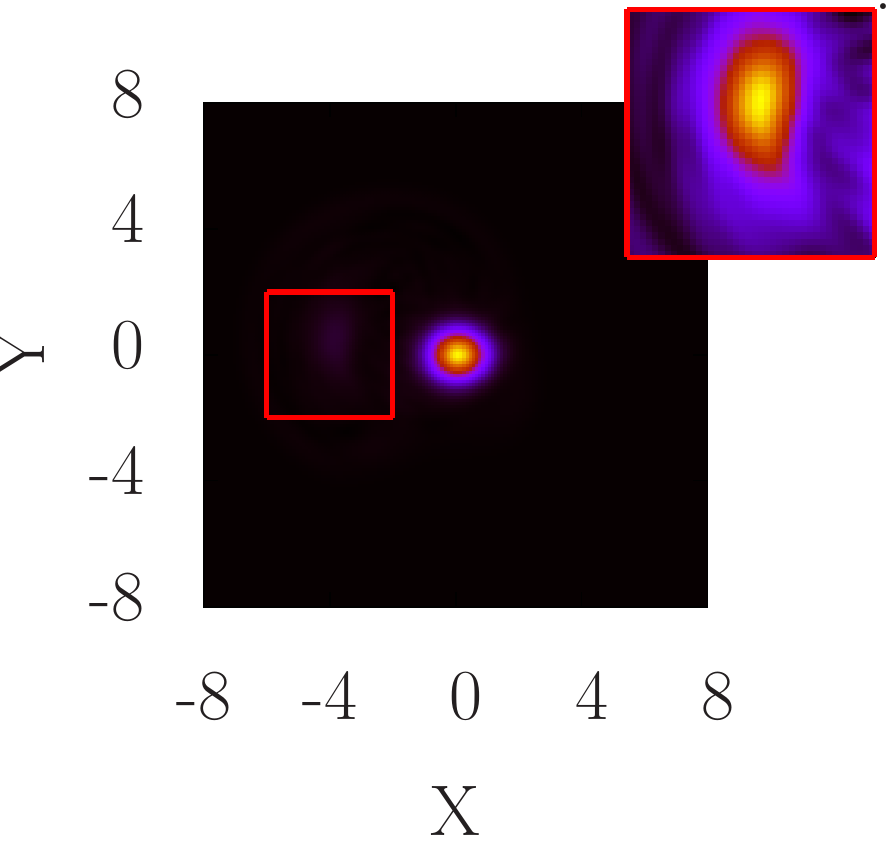} & \includegraphics[width=0.45\linewidth]{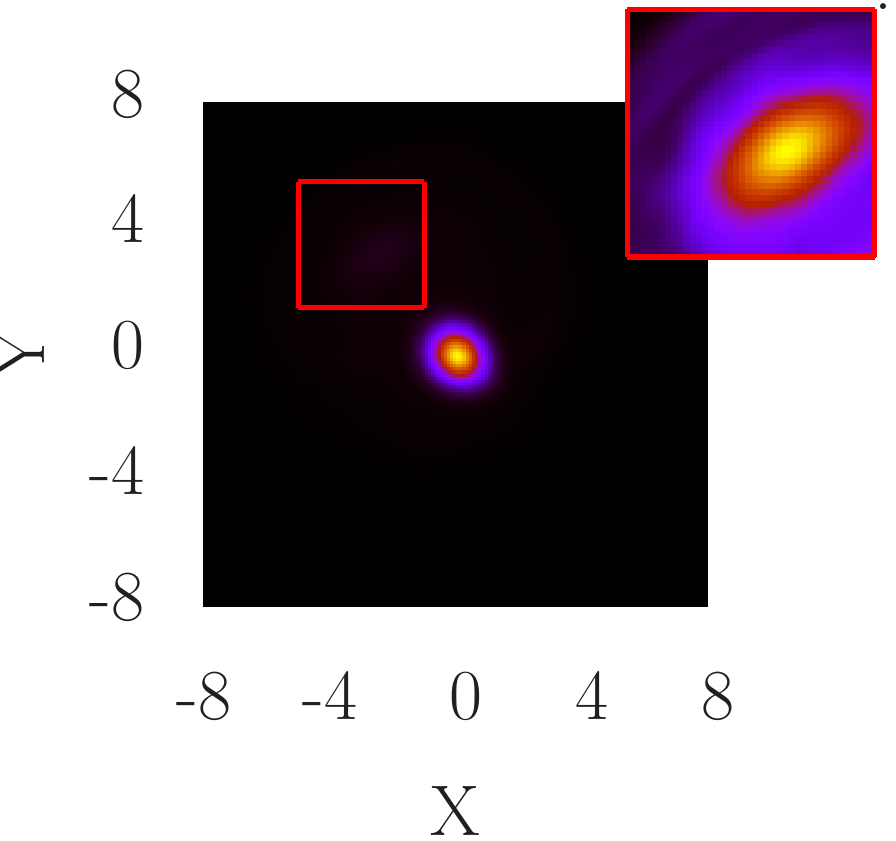}
		\end{tabular}
	\end{center}
	\caption{Wigner functions for the mode in steady-state, with the same parameters as in Fig.~\ref{fig:spectrum_N1}(d). For the plots in (a) the detuning is $\Delta_M=-0.17 g$, indicated by I in  Fig.~\ref{fig:spectrum_N1}(d), which is in the bistability domain. The plots in (b) correspond to neighboring detuning values $\Delta_M + 0.006 g$. The insets zoom into the region indicated by the red box in the main panels, and a different color code scale is used in order to make the peak here visible. In (c) we use the same $\Delta_M$ as for (a), but the phase of the drive amplitude is shifted by $\pi/4$. The quasi-classical component is consequently rotated by the same angle in phase space.}
	\label{fig:WignerBistabN1}
\end{figure}

The full description of the steady state of the cavity mode can be provided in terms of the Wigner quasi-distribution function in phase space. This is displayed in Fig.~\ref{fig:WignerBistabN1}(a) for the detuning indicated by `I' in Fig.~\ref{fig:spectrum_N1}(d).
The Wigner function exhibits two peaks localized both in radial (photon-number distribution) and angular (phase distribution) directions in phase space. Each of these peaks can be considered quasi-classical states. Since there is no negative part of the Wigner function, the reduced density operator of the mode is a mixture of the two quasi-classical states which correspond to the two attractors of the bistability. One of them is the vacuum, the other one is a slightly distorted coherent state. For the given set of parameters, the second peak is centered on the real axis at about -4, that is, this attractor is well separated from the vacuum.  The character of this solution is robust and quite independent of the fine tuning of the driving frequency and strength in a broad range.  This we intend to illustrate with the Wigner function in Fig.~\ref{fig:WignerBistabN1}(b) obtained for  a slightly different detuning. We will show below that the dependence of the state on detuning is smooth in the quantum bistability range, which is at variance with the resonance-like behavior in the MP range. 

The excited quasi-classical component in the mixed steady-state originates from a high-order process. However, unlike the multi-photon resonances, here the excitation at a given drive detuning is not confined into a two-state subspace. For large number of $n$, the neighboring $n$-photon and $(n+1)$-photon transitions to the $|n,-\rangle$ and $|n+1, -\rangle$ states, respectively, are not far in frequency.  The difference is $1/\sqrt{n}-1/\sqrt{n+1} \approx 1/2n^{3/2}$, which vanishes for increasing $n$. Therefore, the highly excited part of the Jaynes--Cummings spectrum is close to an equidistant ladder. The relevant part of the spectrum for this driving frequency can be seen as that of a normal harmonic oscillator with the low-lying steps removed.  More precisely, the low lying number states in the anharmonic part of the spectrum are shifted into non-resonant positions and cannot be populated. Although the intermediate steps are `missing',  the high-lying harmonic part of the spectrum can be excited via high-order processes. This numerically found result is somewhat  surprising since  the given strength of the pump is not enough to induce  multi-photon processes when tuned at resonance to lower states $|n,-\rangle$. The essential difference is that here, in the range `QB', the transitions to the full ladder of the high-lying harmonic part of the spectrum have to be summed up whereas in the range `MP'  the coherent driving is confined into a two-level space. For MP resonances, a spontaneous decay process leads out from the two-state subspace to lower lying levels and the drive cannot re-excite the system, because of the large frequency mismatch, until it reaches the ground state by decay. By contrast, in the QB domain, once the system is prepared in the almost equidistant spectrum part, the decay is continuously balanced by the drive, just like in the case of a driven and lossy harmonic oscillator. 
The calculated Wigner function reveals that these high number-state components form almost a coherent state; the noticeable elongation along the angular direction can be attributed to the residual anharmonicity.  This analogy to coherent driving of a harmonic oscillator is further supported by Fig.~\ref{fig:WignerBistabN1}(c), in which we show that the excited quasi-classical component inherits the phase of the driving  $\eta$. 

\begin{figure}
	\begin{center}
		\includegraphics[width=\columnwidth]{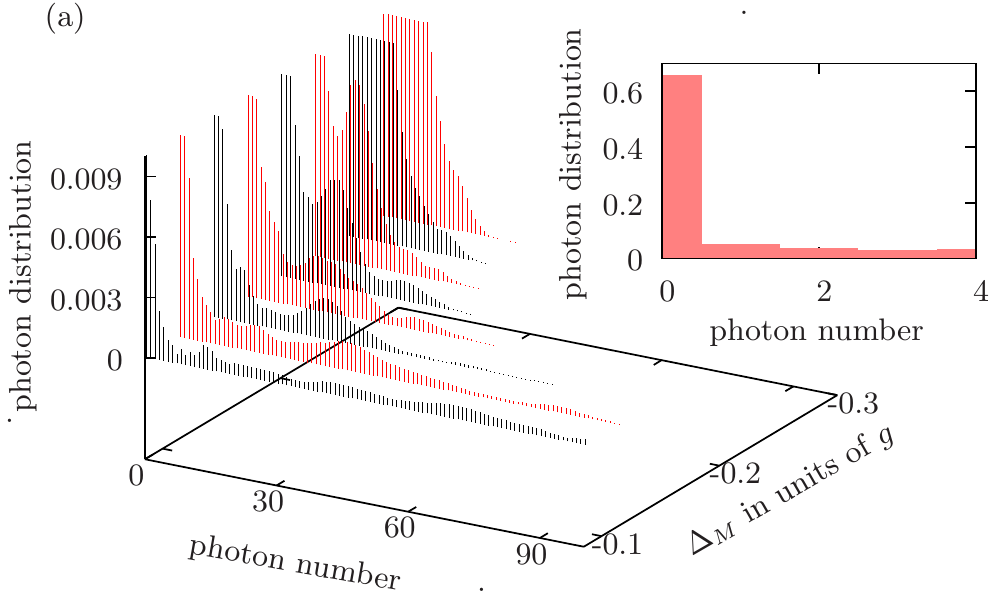} \\
		\includegraphics[width=0.97\columnwidth]{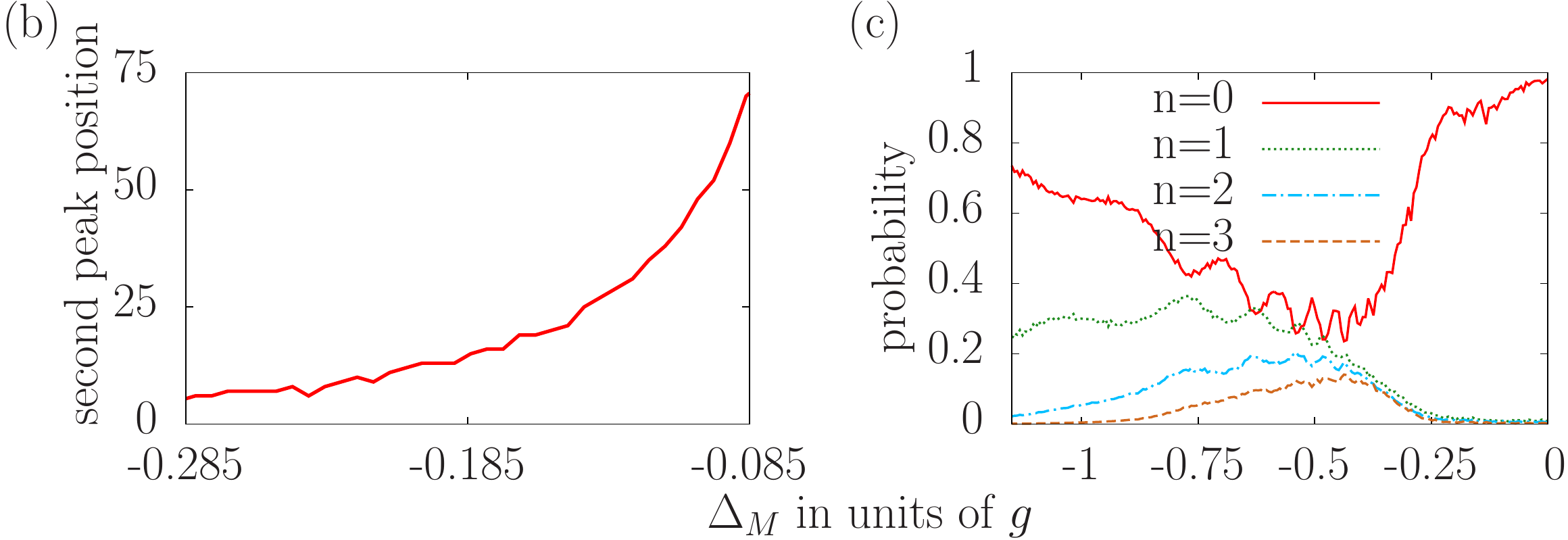}
	\end{center}
	\caption{(a) Steady-state photon number distribution for various detunings in the quantum bistability domain. These distributions are at the origin of the mean photon numbers on the curve associated with $\eta=80 \mu$s${}^{-1}$ in Fig.~\ref{fig:spectrum_N1}. In the main panel the vertical range is adjusted in order to see the large photon number part and thus the distribution for small photon numbers is cut off.  The inset presents the full range of the distribution for $\Delta_M=-0.3 g$, which is single-peaked at vacuum. In (b), the distance of the upper peak of the bimodal distribution from the vacuum is shown to increase nonlinearly as a function of $\Delta_M$. Panel (c) shows the photon number probability for $n=0,1,2,3$ in the full range of the detuning. Depletion dips of the ground state population accompanied by peaks in the $n=1,2,3$ populations reflect the resonance-like multi-photon transitions for large detuning $\abs\Delta_M\gtrsim-0.35 g$. 
In the quantum bistability range near $\Delta_M=0$, the system is mostly in the ground state .}
	\label{fig:PhotonDistribution}
\end{figure}

The steady-state can be characterized by the photon number distribution, since the two peaks in the Wigner function appear along the radial direction. Fig.~\ref{fig:PhotonDistribution} shows how the distribution depends on the detuning $\Delta_M$ in the QB domain. 
The detuning $\Delta_M=-0.3 g$ leads to a photon number distribution which is simply peaked at $n=0$. This detuning corresponds to  the dip between the multi-photon resonance and quantum bistability domains.  When $\abs{\Delta_M}$ is decreased, there appears a secondary peak which corresponds to the excited quasi-classical component revealed by the Wigner function.  The excited quasi-classical component appears only with a small probability,  of course, most of the population remains in the ground state.  Even higher-order peaks can be recognized in the photon number distribution as $\Delta_M$ goes towards $-0.1 g$, that, however, cannot be resolved in the Wigner function representation. 

The photon number corresponding to the center of the secondary peak moves away from $n=0$ towards higher values when the detuning is decreased, as shown in Fig.~\ref{fig:PhotonDistribution}(b). Apart from the numerical noise, this curve is smooth and demonstrates that the bistability solution is not resonance-like and is independent of the fine tuning of the driving field.

The total population in the photon number states belonging to the secondary peak decreases. This can be ascertained indirectly, from the plot Fig.~\ref{fig:PhotonDistribution}(c) showing the population in the zero photon state $\ket{0}$. This population is monotonously growing in the detuning range out of the multi-photon transitions ($\abs\Delta_M\lesssim-0.35 g$, disregarding the small wiggles of numerical origin).  The decreasing weight and the increasing photon number of the secondary peak are the two competing tendencies which govern the variation of the mean photon number. It is interesting that the combination of small photon numbers with large weight and larger photon numbers with small weight in the mixed steady-state density matrix amounts to the same mean photon number as the one derived from the \emph{ab initio} semiclassical theory on the rising slope of the peak, cf. the agreement with the semiclassical dashed blue curve in the range $-0.2 g\lesssim \Delta_M \lesssim -0.15 g$ in Figs.~\ref{fig:spectrum_N1}. Note that the semiclassical solution in this range is simply that of an driven cavity without atoms. Although the mean field description accidentally yields a correct mean intensity in this limited range of detuning, it does not account for the highly non-trivial photon number distribution.  

\begin{figure}
	\begin{center}
			\includegraphics[width=\linewidth]{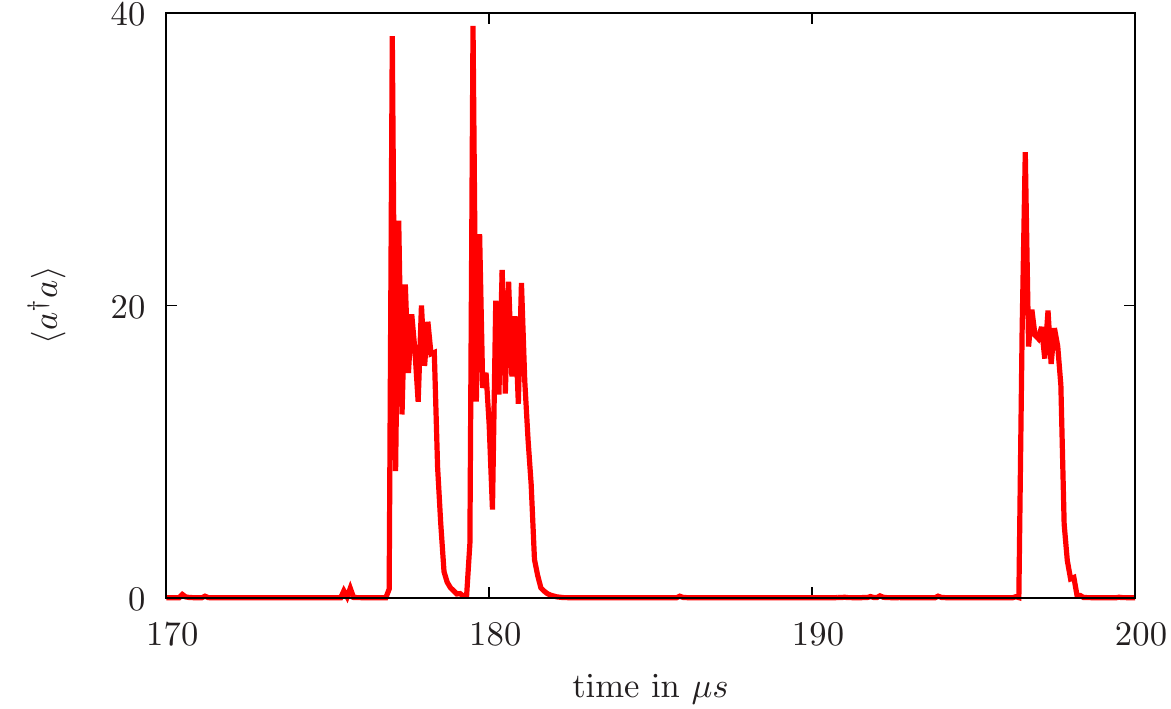} \\
	\end{center}
	\caption{Fluctuations of the instantaneous quantum average of the photon number operator along simulated quantum trajectories  during a period of $30\mu$s. The telegraph signal is the temporal behavior in the bistability domain. Parameters are the same as in Fig. \ref{fig:spectrum_N1}(d), the given value of detuning is marked by the vertical dashed line on those figures as I, $\Delta_M=-0.17$ in units of $g$.}
	\label{fig:Trajectories}
\end{figure}

The bistability associated with such a two-peaked Wig\-ner function can be easily seized in the temporal behaviour. We plot the time evolution of the instantaneous quantum average of the photon number operator along the simulated quantum state trajectory. This is presented in Fig.~\ref{fig:Trajectories}.
An arbitrary period of 30$\mu$s long after the initial transient is displayed for the detuning marked by I in \ref{fig:spectrum_N1}(d). The quantum trajectory of the average intensity shows the alternation of periods with finite photon number around $\langle a^\dagger a\rangle \approx 20$ and periods where the state is close to the ground state. The evolution is reminiscent of a telegraph signal exhibiting a long time scale associated with the switching time between the two attractors. One can distinguish the fluctuations induced by quantum jumps, i.e., the small wiggling of the mean intensity plateau at about $\langle a^\dagger a\rangle \approx 20$, from the sharp upsurges associated with switching. Fluctuations are due to the dissipative processes, whereas the switching originates from a highly nonlinear dynamics.

The emergence of the characteristic switching time explains the vanishing of the mean photon number in Figs.~\ref{fig:spectrum_N1} when the detuning reaches the close vicinity of $\Delta_M=0$.  The systematic variation of the steady-state photon number distribution with the two peaks as described above in connection with Fig.~\ref{fig:PhotonDistribution}, is not expected to change at this detuning range. However, as the secondary peak moves away and the separation of the attractors increases, the switching time diverges and the steady state cannot be reached within the finite duration of the numerical simulation. This statistical effect explains also why the quantum bistability peak in Fig.~\ref{fig:spectrum_N1} is quite noisy: it takes very long time to accumulate a sufficiently good statistics.  The suppression of the mean photon number at $\Delta_M=0$ is thus simply a finite-time effect. It has a remarkable consequence: the bistability feature at exact resonance $\Delta_M=\Delta_A=0$, i.e., absorptive bistability  cannot be observed in reasonable time.

\section{Conclusions}
\label{sec:Conclusions}

Circuit QED setups represent systems reaching hitherto unexplored regimes of the Jaynes-Cummings model. The qubit-photon coupling is so strong that the saturation photon number is tiny. The so-called optical bistability effect, which relies on the nonlinearity of the atom-light interaction, vanishes and there is no quantum limit of this effect for such a fractional saturation photon number. On one hand, the nonlinear response of the coupled qubit-photon system can directly reflect the anharmonic spectrum of the driven Jaynes-Cummings model. This response is dramatically nonlinear, since it involves multi-photon resonances. On the other hand, we identified a detuning range without multi-photon resonances in which the quantum system evolves into a bistability-like steady-state. The system sporadically flips between the ground state and a highly excited quasi-classical state with well-defined phase and amplitude. This solution with amplitude bimodality, obtained at finite detuning between the drive and the cavity mode frequencies, is markedly different from the spontaneous dressed-state polarization that has been predicted for the resonant case and $\gamma\ll\kappa$ \cite{alsing1991spontaneous,soklakov2001conditional}. Further work is needed to clarify the relation of these two solutions, e.g., to investigate the effect of gradually increasing $\gamma$ from zero. In the present work we performed numerical simulations at the borders of current possibilities. In other words, these phenomena are at the limit where the deeper insight into Jaynes-Cummings physics calls for experimental approach and veritable quantum simulation.

\section*{Acknowledgements}

This work was supported by the EU FP7 (ITN, CCQED-264666), the Hungarian National Office for Research and Technology under the contract ERC\_HU\_09 OPTOMECH, and the Hungarian Academy of Sciences (Lend\"ulet Program, LP2011-016). We acknowledge NIIF for awarding us access to resource based in Hungary at P\'{e}cs. A. V. acknowledges support from the J\'anos Bolyai Research Scholarship of the Hungarian Academy of Sciences.

\end{document}